\documentclass[12pt,preprint,english,A4]{iopart}
\usepackage[dvips]{graphicx}

\usepackage{times}

\newcommand{\Ms}{M_{\star}}

\newcommand{\Rs}{R_{\star}}
\newcommand{\Mbh}{M_{\bullet}}
\newcommand{\Mo}{M_{\odot}}

\newcommand{\kmpers}{{\rm km\,s^{-1}}}
\newcommand{\pc}{{\rm pc}\,}
\newcommand{\mpc}{{\rm mpc}\,}

\newcommand{\Gyr}{{\rm Gyr}\,}

\newcommand{\beq}{\begin{equation}}
\newcommand{\eeq}{\end{equation}}

\newcommand{\aj}{{\it Astron. J.}}
\newcommand{\apj}{{\it Astrophys. J.}}
\newcommand{\apjl}{{\it Astrophys. J. Lett.}}
\newcommand{\mnras}{{\it Mon. Not. R. Astron. Soc.}}
\newcommand{\prd}{{\it Phys. Rev. D.}}
\newcommand{\physrep}{{\it Phys. Rep.}}
\newcommand{\nat}{{\it Nature}}

\def\apgt{\ {\raise-.5ex\hbox{$\buildrel>\over\sim$}}\ } \def\aplt{\
{\raise-.5ex\hbox{$\buildrel<\over\sim$}}\ }

\begin{document}

\title[Extreme mass ratio inspiral rates]{Extreme mass ratio inspiral rates: dependence on the massive black hole mass}

\author{Clovis Hopman} 

\address{Leiden University, Leiden Observatory,
P.O. Box 9513, NL-2300 RA Leiden}

\begin{abstract}
We study the rate at which stars spiral into a massive black hole
(MBH) due to the emission of gravitational waves (GWs), as a function
of the mass $\Mbh$ of the MBH. In the context of our model, it is
shown analytically that the rate approximately depends on the MBH mass
as $M_{\bullet}^{-1/4}$. Numerical simulations confirm this result,
and show that for all MBH masses, the event rate is highest for
stellar black holes, followed by white dwarfs, and lowest for neutron
stars. The Laser Interferometer Space Antenna (LISA) is expected to
see hundreds of these extreme mass ratio inspirals per year. Since the
event rate derived here formally diverges as $\Mbh\!\to\!0$, the model
presented here cannot hold for MBHs of masses that are too low, and we
discuss what the limitations of the model are.
\end{abstract}

\section{Introduction}

The inspiral of a compact remnant into a massive black hole (MBH)
leads to the emission of gravitational waves (GWs) with frequencies
that will be detectable by the Laser Interferometer Space Antenna
(LISA). Such extreme mass ratio inspirals (EMRIs) will lead to a
wealth of astrophysical information, such as highly precise
measurements of the mass and spin of the MBH, the mass of the
inspiraling object, the distance to the source, and the direction of
the source on the sky \cite{Bar04a}. LISA will be sensitive to MBHs
with masses in the range $10^4\Mo\aplt\Mbh\aplt10^7\Mo$, and is
expected to detect thousands of events during its mission lifetime
\cite{Gai04}. The existence of MBHs with masses $\Mbh<10^6\Mo$ is
still hypothetical, in spite of considerable circumstantial evidence
(see \cite{Mil04b} for a review). Some of this
evidence stems from ultra-luminous X-ray sources (ULXs). These are
sources of X-rays that are more luminous than the Eddington luminosity
for a $10\Mo$ stellar black hole (BH), but are located away from the
center of their host galaxy, which makes it improbable that they are
very massive, as dynamical friction would have brought them to the
center. Arguments that selected ULXs are more massive than stellar BHs
include mass estimates from the mass-temperature relation
\cite{Mil03, MilFabMil04} and quasi-periodic oscillations
\cite{Fio04, Liu05}. Dynamical analysis also indicates that some
globular clusters may contain quiescent intermediate mass BHs
\cite{Ger02, Geb05, Noy08}.

The rate at which EMRIs form near a given MBH is still rather
uncertain, in part due to the scope of different dynamical processes
that need to be taken into account. Estimates of event rates
\cite{Hil95, Sig97b, Mir00, Fre01, Iva02, Fre03, Ale03b, Hop05, Hop06,
Hop06b} have typically focused on MBHs with masses of
$\Mbh\sim10^6\Mo$ (see \cite{Hop06c} for a review). One reason for
this is that the signal for such a MBH is strongest, but another, more
practical reason, is that in that case the system can be scaled with
the situation in the Galactic center, which contains a MBH of
$\Mbh\approx(3-4)\times10^6\Mo$ \cite{Ghe98, Sch02, Eis05, Ghe08}. The
Galactic center has been studied in great detail (see \cite{Ale05} for
a review), and the knowledge we have of the stellar properties of the
Galactic center can be exploited to calibrate models of stellar
galactic nuclei.

A possible Ansatz to estimate event rates for lower mass MBHs is to
assume that the empirical $\Mbh-\sigma$ relation 
\begin{equation}\label{e:ms}
\Mbh = 10^8\Mo\left({\sigma\over200\kmpers}\right)^{4}
\end{equation}
between the $\Mbh$ and the velocity dispersion of the host galaxy
\cite{Fer00, Geb00, Tre02} extends to lower mass MBHs. By making this
assumption, inspiral rates for these hypothetical MBHs can be
estimated. It should be stressed that the continuation of the
$\Mbh-\sigma$ relation to lower masses has no empirical foundation -
indeed, the very existence of lower mass MBHs is not universally
accepted.

One obvious interest in the detection of the inspiral into a lower
mass MBH by LISA is that it would immediately provide unambiguous
proof for the existence of such MBHs. In addition, an interesting
possibility that is unique for lower mass MBH was suggested by
\cite{Men08, Ses08}. Since the tidal radius
$r_t=\left(\Mbh/\Ms\right)^{1/3}\Rs$ of a star with mass $\Ms$ and
radius $\Rs$ has a weaker dependence on $\Mbh$ than the Schwarzschild
radius $r_S=2G\Mbh/c^2$, tidal effects become more important compared
to general relativistic effects as $\Mbh$ decreases. As a result, when
MBHs of mass $\Mbh\aplt10^5$ capture a white dwarf (WD), the WD may
be disrupted by the tidal field during the last phases of
inspiral. This may lead to a very interesting coincidence of a LISA
detection with an electromagnetic counterpart. Lower mass MBHs also
contribute differently to the stochastic background of GWs from EMRIs,
due to the GWs emitted prior to the phase where they reach a signal to
noise ratio high enough for detection \cite{Bar04b}.

These questions motivate us to revisit the dependence of the inspiral
rate $\Gamma$ on $\Mbh$. \cite{Hop05} used analytical arguments to
find $\Gamma\propto\Mbh^{-1/4}$. Here we rephrase and extend these
arguments for systems that include mass-segregation
(\S\ref{s:model}). We then perform a series of numerical simulations,
to derive the dependence $\Gamma(\Mbh)$ numerically
(\S\ref{s:results}), and show that the dependence is in good agreement
with the analytical result. We discuss the validity, extensions, and
consequences of our model in (\S\ref{s:disc}).

\section{EMRI rate model and scaling with mass}\label{s:model}

In order to compensate for our ignorance about lower mass MBH systems,
we make the Ansatz discussed in the introduction of assuming that the
$\Mbh-\sigma$ relation can be extended to $\Mbh<10^6\Mo$, and we
assume that the mass enclosed within the radius of influence
$r_h=G\Mbh/\sigma^2$ is proportional (and comparable) to that of the
MBH. Based on these assumptions, we derive an analytical expression
(\ref{e:GamEMRIstep}) for the dependence of the event rate on $\Mbh$
and present a numerical model to calculate these event rates.

\subsection{Scaling galactic nuclei with mass}\label{ss:scaling}

With the aforementioned assumptions, we find a number of scaling
relations. The radius of influence, where the MBH dominates the
potential, is given by

\begin{eqnarray}\label{e:rh}
r_h &=& {G\Mbh\over\sigma^2}\nonumber\\
&=&2\pc \left({\Mbh\over 3\times10^6\Mo}\right)^{1/2}.
\end{eqnarray}

The fact that the enclosed mass of stars within $r_h$ is proportional
to the MBH mass implies that the number density at the radius depends
on $\Mbh$ as

\begin{eqnarray}
n_h &=&4\times10^4\pc^{-3}\left({\Mbh\over 3\times10^6\Mo}\right)^{-1/2}
\end{eqnarray}
(for the normalization, see \cite{Gen03a}).

The relaxation time then increases with the mass of the MBH as

\begin{eqnarray}\label{e:Th}
T_h &=& {3(2\pi\sigma^2)^{3/2}\over 32\pi^2 G^2 \Ms^2 n_h\ln\Lambda}\nonumber\\
    &=& 6\,\Gyr\left({\Mbh\over 3\times10^6\Mo}\right)^{5/4}.
\end{eqnarray}
In this expression, the Coulomb logarithm is approximately
$\ln\Lambda=\ln\left(\Mbh/\Ms\right)$; for our analytical arguments we
discard the weak mass-dependence of the Coulomb logarithm.

\subsection{Model for EMRI rate}\label{ss:rate}

We follow the analysis presented in \cite{Hop06b}, who solve the
Fokker-Planck equations for a multi-mass model with 4 species: main
sequence stars (MSs) of mass $M_{\rm MS}=\Mo$; WDs of mass $M_{\rm
WD}=0.6\Mo$; neutron stars (NSs) of mass $M_{\rm NS}=1.4\Mo$; and
stellar BHs of mass $M_{\rm BH}=10\Mo$. In all cases we assume that
the stars far away from the MBH have equal velocity dispersions
$\sigma$, and that the unbound stars have a distribution function
\begin{equation}\label{e:BC}
g_M(x)=C_M\exp(x)\quad\quad(x<0),
\end{equation}
where $x=E/\sigma^2$ is the dimensionless energy of the stars, and $E$
is the negative specific energy with respect to the MBH. The prefactor $C_M$ determines the
fraction of compact remnants of type $M$ in the unbound stellar
population. The distribution (\ref{e:BC}) is what may be expected if the 
system was formed by violent relaxation; the stellar distribution of stars
bound to the MBH is not very strongly affected by the exact form of this 
distribution (although it is dependent on its normalization), see e.g. 
\cite{Ale08}. To simplify our analysis, we do not consider the evolution
of unbound stars, and assume that they are in steady state. This is
mainly a simplification, but there is some justification in the fact
that the relaxation time far away from the MBH increases rapidly with
distance, so that the dynamical evolution of these stars is
slow. Also, there is likely star formation far away from the MBH,
which will be harder to model, so that it is not clear how much
realism would be obtained by including dynamics far away from the
MBH. We note that the stellar models presented in \cite{Fre01, Fre06}
do include dynamics for all stars in the system. For our models, we
will assume that $C_M = (1, 0.1, 0.01, 0.001)$ for MSs, WDs, NSs and
BHs, consistent with models of continuous star formation that appear
to be consistent with Galactic center observations \cite{Ale99b,
Ale05}.

The Fokker-Planck equations include a loss-cone term, and in
dimensionless units, the equations have the following form (for
details, see \cite{Hop06b}; \cite{Ale08})

\begin{equation}\label{e:dgmdt}
{\partial g_M(x, \tau)\over\partial\tau} =  -x^{5/2} {\partial \over \partial x}Q_{M} - R_{M}(x),
\end{equation}
where $R_{M}(x)$ is the rate at which stars of type $M$, with energies
in the interval $(x, x+dx)$, reach the loss-cone and are eaten by the
MBH. The rate at which stars flow through a given energy $x$ is given
by $Q=Q[g(x, \tau),x]$; this non-linear term represents 2-body
relaxation. We stress that these equations do not depend on the MBH
mass (apart from the assumption that it is much more massive than any
of the individual stars). The same Fokker-Planck equation therefore
applies to all systems, regardless of their MBH. The mass does enter,
of course, in the dimension-full equations, and dictates for example
how many years it takes to reach steady state.

The rate at which stars enter the loss-cone is much higher than the
EMRI formation rate: many stars plunge prematurely without spiraling
in. These stars never reach an orbit detectable by LISA.\footnote{An
exception to this is the Galactic center where gravitational bursts
from fly-bys are possibly detectable \cite{Rub06, Hop07, Yun08}.}

In dimension-full units, the rate at which stars with semi-major axes
less than $a$ come into the loss-cone is then

\begin{equation}\label{e:Gam}
\Gamma_{M}(<a) = 2\sqrt{2}I_0\int_{0}^{a/r_h}dyy^{1/2}R_{M}\left(y\right),
\end{equation}
where the flow $I_0$ is given by

\begin{eqnarray}
I_{0}&=&{\frac{8\pi^{2}}{3\sqrt{2}}}r_{h}^{3}n_h{\frac{(G\Ms)^{2}\ln\Lambda n_h}{\sigma^{3}}}\nonumber\\
&=&10^{5}\,\Gyr^{-1}\left({\Mbh\over 3\times10^6\Mo}\right)^{-1/4},
\end{eqnarray}
which is approximately the number of the stars within $r_h$ divided by
the relaxation time. The actual flow is in some situations much
smaller than $I_0$ (i.e.,
$\int_{0}^{a/r_h}dyy^{1/2}R_{M}\left(y\right)\ll1$), as was realized
by \cite{Bah77}, who derived several analytical results from such
``zero-flow'' solutions. However, it was shown by \cite{Ale08} that in
cases such as here, where there are very rare, massive objects (i.e.,
stellar BHs), maximal flows can in fact be achieved.

For LISA, the only objects that are accreted by the MBH and detected are those
(compact) stars that spiral into the MBH gradually, without plunging
prematurely into the loss-cone. Therefore, for in-spirals, the
integrand in equation (\ref{e:Gam}) should be convolved with a
function $S_{M}(a)$ that gives the probability for a star to spiral in
without plunging. This function filters out all stars that plunge into 
the MBH without leading to a detectable EMRI source ("plunges"), and 
selects only the stars that do lead to a slowly inspiraling star. We 
compute this function with the Monte Carlo methods as described in 
\cite{Hop05} (see also \cite{Hil95}, who used a very similar method). 
The total EMRI rate is then given by


\begin{equation}\label{e:GamEMRI}
\Gamma_{M}(<a) = 2\sqrt{2}I_0\int_{0}^{a/r_h}dy y^{1/2}S_M(y)R_{M}\left(y\right).
\end{equation}
The function $S_M(a/r_h)$ selects only those captured stars that
spiral in successfully, which is a small fraction of the total number
of captures [so
$\int_{0}^{a/r_h}dyy^{1/2}S_M(y)R_{M}\left(y\right)\ll\int_{0}^{a/r_h}dyy^{1/2}R_{M}\left(y\right)$],
of order 10\% \cite{Ale03b}.

In contrast to the solution of the dimensionless Fokker-Planck
equations, the function $S_M(a)$ may depend on the MBH mass, because
the loss-cone, which is determined by the angular momentum of the last
stable orbit $J_{\rm LSO}=4G\Mbh/c$, depends on $\Mbh$, and provide an
additional length scale to the problem. However, it was argued by
\cite{Hop05}, that $S_M(a)$ can be approximated by a step-function
that is zero for $a>a_c$, with

\begin{equation}
\frac{a_{c}}{r_{h}}=\left(\frac{d_{c}}{r_{h}}\right)^{3/(3-2p_{M})},\label{e:ac}\end{equation}
where
\begin{equation}
d_{c}\equiv\left(\frac{8\sqrt{G\Mbh}E_{1}T_{h}}{\pi c^{2}}\right)^{2/3};\quad\quad E_{1}\equiv\frac{85\pi}{3\!\times\!2^{13}}\frac{\Ms c^{2}}{\Mbh},\label{e:D}
\end{equation} 
and $p_{M}$ is the slope of the distribution function $g_{M}(x)$. It
follows that $d_c\propto\Mbh^{1/2}$, and hence that $a_c/r_h$ is
independent of $\Mbh$ (see also figure \ref{f:SaWD}). Hence, at the
level of this approximation, it follows that the integral in equation
(\ref{e:GamEMRI}) does not depend on mass, and thus that

\begin{eqnarray}\label{e:GamEMRIstep}
\Gamma_{M} &=& 2\sqrt{2}I_0\int_{0}^{\infty}dy y^{1/2}S_M(y)R_{M}\left(y\right)\nonumber\\
&\approx& 2\sqrt{2}I_0\int_{0}^{a_c/r_h}dy y^{1/2}R_{M}\left(y\right)\nonumber\\
&\propto&I_0\propto\Mbh^{-1/4}.
\end{eqnarray}

The result (\ref{e:GamEMRIstep}) is derived from the assumption that
$S_{M}(y)$ may be approximated by a step function
$S_M(y)\approx\theta\left(a_c/r_h-y\right)$ with $a_c/r_h$ as in
equation (\ref{e:ac}). This result is based on several approximations
compared to the numerical Fokker-Planck / Monte Carlo model. To test
the validity of this approximate expression, we perform Monte Carlo
experiments to determine the function $S_{M}$.

The calculation for the EMRI rate for all systems we consider here
then goes through the following steps. (i) calculate the steady state
of equations (\ref{e:dgmdt}) once for all systems. (ii) From the
outcome, find the dimensionless relaxation time

\begin{equation}
\tau_r(x) = {\Mo^{2}\over\sum_{M}g_M(x)M^{2}}.
\end{equation}
(iii) Convert this dimensionless time-scale into a dimension-full time
through equation (\ref{e:Th}), and use this time in the Monte Carlo
simulations to find the function $S_{M}(a)$. (iv) Calculate the
dimension-full integral (\ref{e:GamEMRI}) to calculate the EMRI rate.

\section{Numerical results}\label{s:results}
We consider MBH masses of $\Mbh=3\times10^4\Mo$,
$\Mbh=3\times10^5\Mo$, and $\Mbh=3\times10^6\Mo$, with stellar
populations as indicated in \S\ref{s:model}.

\begin{figure}[!h]
\includegraphics[height=130 mm,angle=270 ]{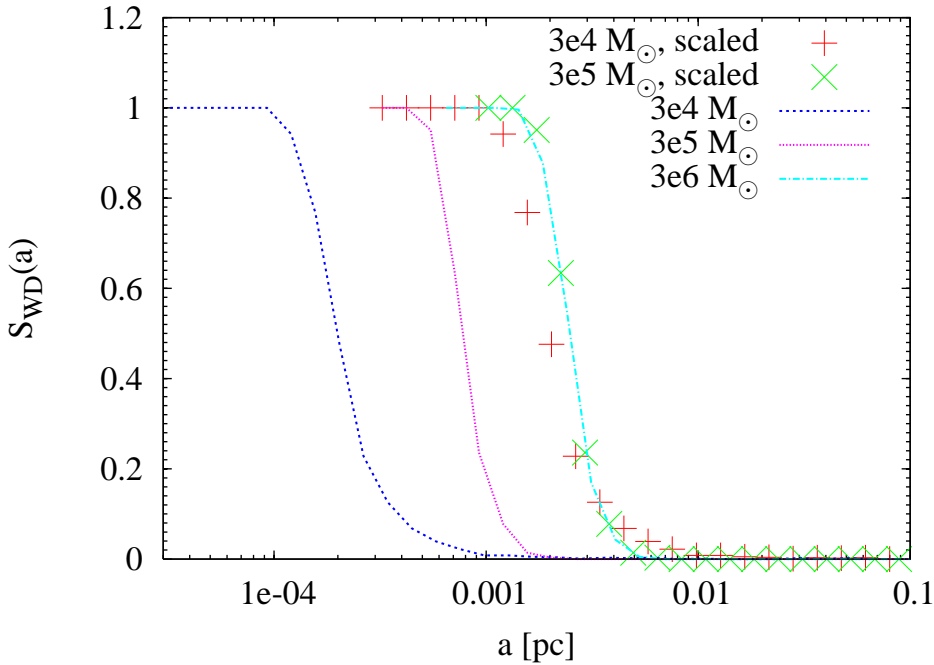}
\caption{Inspiral fraction $S_{\rm WD}(a)$ for WDs, for several MBH
masses. The solid lines show the direct results, whereas the crossed
data points were re-scaled by $\sqrt{3\times10^6\Mo/\Mbh}$.  The
inspiral rate is high for small semi-major axes, and drops to zero at
large semi-major axes, with $S_{\rm WD}=0.5$ at $\sim
2\mpc\,\left(\Mbh/3\times10^6\Mo\right)^{-1/2}$. The fact that the
scaled lines coincide with the line for $\Mbh=3\times10^6\Mo$ implies
that $S_M(a/r_h)$ is to good approximation independent on the MBH
mass. \label{f:SaWD}}
\end{figure}

In figure (\ref{f:SaWD}), we show the inspiral fractions $S_{\rm
WD}(a)$ for WDs for the MBH masses considered. Inspiral probabilities
are high for small semi-major axes, as anticipated. The analytical
arguments given in the previous section suggest that $S_{\rm
WD}(a/r_h)$ does not strongly depend on the MBH mass. To test this
idea, we multiply the horizontal axis by $\sqrt{3\times10^6\Mo/\Mbh}$
in order to extract the MBH dependence of $r_h$ (equation
\ref{e:rh}). The curves now coincide approximately, confirming that
the dependence on $\Mbh$ is weak. The figure also shows that $S_{\rm
WD}(a)$ can be reasonably approximated by a step-function.

The resulting event rates are reported for all different species in
table (\ref{t:t1}). The numbers for $\Mbh=3\times10^6$ are different
by a factor $\aplt2$ compared to those in \cite{Hop06b}, probably
mainly due to slightly different boundary conditions.

\begin{figure}[!h]
\begin{center}
\includegraphics[height=130 mm,angle=270 ]{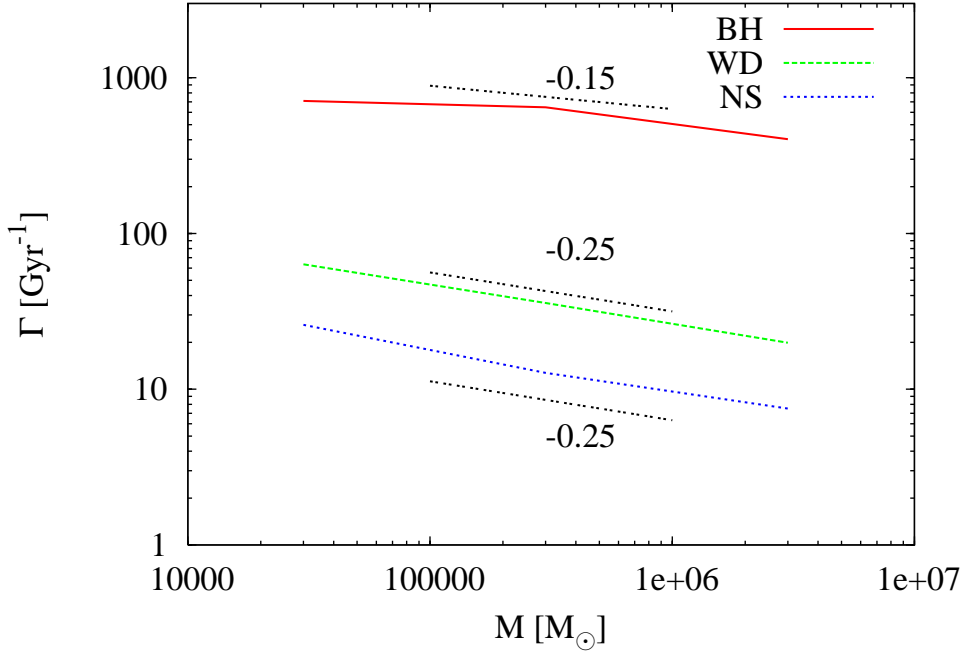}
\caption{EMRI event rates as a function of MBH mass for BHs (top
line), WDs (middle line), and NSs (bottom line). The straight lines
give approximate power laws. For WDs and NSs the slope of $-0.25$ is
in very good agreement with the analytical result, while for BHs, the
slope of $-0.15$ is slightly flatter.\label{f:Rates}}
\end{center}
\end{figure}

The event rates are plotted for the different species as a function of
mass in figure (\ref{f:Rates}). This figure shows that the event
rate indeed decreases with $\Mbh$, and the relation $\Gamma_M\propto
\Mbh^{-1/4}$ is confirmed.

\begin{table}[t]
\caption{EMRI rates}\label{t:model}
\begin{tabular}{llll}
\hline
\hline
$M$            & WD            &   NS            &     BH           \tabularnewline
$[\Mo]$        & $[\Gyr^{-1}]$ & $[\Gyr^{-1}]$   &   $[\Gyr^{-1}]$  \tabularnewline
\hline
$3\times10^4$  & 62            &   26            &    710          \tabularnewline
$3\times10^5$  & 34            &   13            &    650           \tabularnewline
$3\times10^6$  & 20            &   7             &    400           \tabularnewline
\hline
\label{t:t1}
\end{tabular}
\end{table}

\section{Summary and discussion}\label{s:disc}

We analyzed the EMRI rate for several MBH masses in the context of the
\cite{Hop05} and \cite{Hop06b} models. The rate depends on MBH mass
as $\Gamma_M\propto \Mbh^{-1/4}$. The analytical mass dependence
(equation \ref{e:GamEMRIstep}) is confirmed by Monte Carlo
simulations.

It is not trivial to calculate the projected EMRI detection rate from
a given event rate \cite{Gai04}. The inspiral rate estimated here
would give rise to a detection rate of $\sim10^3$ over the LISA life
time of 5 years \cite{Gai08}.

Since the event rate increases as the $\Mbh$ mass decreases, the
assumptions for the models here must break down at some point, since
otherwise the MBH would accrete more than its own mass within the age
of the system. This is already the case for the $\Mbh=3\times10^4\Mo$
considered here. Several possible solutions to this paradox are that
(i) such MBHs do not exist, or exist only during transient phases much
shorter than the age of the Universe (for example, {\it because} of
such high capture rates); or that (ii) somehow the assumed boundary
conditions are not fulfilled, for example because the $\Mbh-\sigma$
relation is not satisfied MBHs with $\Mbh\aplt10^6\Mo$, or BHs do
not sink in effectively from large distances, and are not formed by
new star formation, so that the boundary condition for the unbound
stars is not maintained over time. In the latter situation, the WD and
NS rates, which are lower (and bring in less mass), could still be
correct, and perhaps the actual rates would be even higher if stellar
BHs do not push these stars out by mass segregation. To test this
idea, we performed a simulation with a $\Mbh=3\times10^4\Mo$ MBH,
similar to the ones presented in the previous sections, but without
any stellar BHs. In this example, the event rate increases by a factor
$\sim2.5$ to $\Gamma_{\rm WD}=146\Gyr^{-1}$. We finally note that we
find that the MBH grows as $\Mbh\propto t^{4/5}$ as a result of EMRI 
captures, much slower than the exponential growth that would result from
Eddington limited accretion.

In a recent paper, \cite{OLe08} study the stellar BH distribution
around MBHs in a different context, where they consider cases with a
much flatter mass function than we assume here, and with more massive
stellar BHs. This will likely increase the EMRI rate further (and
hence restrict further the MBH range for which the analysis is
valid). Here we do not consider this situation.

The models presented here do {\it not} include the effects of resonant
relaxation \cite{Rau96}. Resonant relaxation increases the rate at
which the angular momenta of the stars randomize, but the effect is
largely destroyed for very eccentric orbits due to general
relativistic precession. It was argued by \cite{Hop06} that resonant
relaxation will likely increase the EMRI rate by a factor of
$\aplt10$. Many aspects of resonant relaxation remain unclear,
however. Although the eccentricity dependence of random torques in a
power-law cusp was found by \cite{Gur07}, the coherence time of the
torque, and its dependence on eccentricity, remain unknown. In
addition, there is the related effect of resonant friction
\cite{Rau96}, which has not been considered so far in the literature
in detail. It can be concluded that resonant relaxation will probable
modify the EMRI event rate, but a more detailed analysis is premature
at this point.

It has been suggested that triaxiality can also affect the EMRI rate 
\cite{Hol06}. Angular momentum is not a conserved quantity in 
a triaxial potential, and as a result it can potentially change much 
faster than can be achieved by relaxation mechanisms. However, it 
seems unlikely that the potential is significantly triaxial within 
the radius of influence, since the potential is there dominated by 
the MBH. Since $S_M(a)=0$ for $a>10^{-2}\pc$ (see figure \ref{f:SaWD}), 
enhanced evolution of angular momenta outside this distance would not lead 
to any additional EMRIs: if more stars are captured by the MBH, these 
captures will all result in plunges. Triaxiality outside the radius of influence 
can  therefore only increase the capture rate of plunges. Similar 
conclusions apply to massive perturbers \cite{Per07, Zha09}.

We have focused on the direct capture of compact remnants by MBHs. An
alternative route to making a LISA source with an intermediate mass
black hole is by tidal capture of a main sequence star which
circularizes near the MBH, and after it leaves the main sequence
spirals into the MBH as a compact remnant \cite{Hop04, Hop05b}. Such
different formation mechanisms may lead to higher event rates.

\ack{I thank Alberto Lobo and Carlos Sopuerta for organizing a very
fruitful LISA symposium. The work presented here was further developed
at the 2W@AEI meeting at the Max-Planck Institut f\"ur
Gravitationsphysik (Albert Einstein-Institut), organized by Pau
Amaro-Seoane. I thank Jonathan Gair, Alberto Sesana and Pau
Amaro-Seoane for encouraging me to investigate the mass dependence of
EMRI rates in more detail, and Tal Alexander and Ann-Marie Madigan 
for comments on the manuscript. This work was supported by a Veni 
fellowship from the Netherlands Organization for Scientific Research (NWO).}


\Bibliography{99}
\expandafter\ifx\csname natexlab\endcsname\relax\def\natexlab#1{#1}\fi

\bibitem{Ale05}
{Alexander}, T. 2005, \physrep, 419, 65

\bibitem
{Ale03b}
{Alexander}, T., \& {Hopman}, C. 2003, \apjl, 590, L29

\bibitem
{Ale08}
---. 2008, ArXiv e-prints, (arXiv: 0808.3150)

\bibitem
{Ale99b}
{Alexander}, T., \& {Sternberg}, A. 1999, \apj, 520, 137

\bibitem
{Bah77}
{Bahcall}, J.~N., \& {Wolf}, R.~A. 1977, \apj, 216, 883

\bibitem{Bar04b}
{Barack}, L., \& {Cutler}, C. 2004{\natexlab{a}}, \prd, 70, 122002

\bibitem{Bar04a}
---. 2004{\natexlab{b}}, \prd, 69, 082005

\bibitem
{Eis05}
{Eisenhauer}, F., {et~al.} 2005, \apj, 628, 246

\bibitem
{Fer00}
{Ferrarese}, L., \& {Merritt}, D. 2000, \apjl, 539, L9

\bibitem
{Fio04}
{Fiorito}, R., \& {Titarchuk}, L. 2004, \apjl, 614, L113

\bibitem
{Fre01}
{Freitag}, M. 2001, Classical and Quantum Gravity, 18, 4033

\bibitem
{Fre03}
---. 2003, \apjl, 583, L21

\bibitem
{Fre06}
{Freitag}, M., {Amaro-Seoane}, P., \& {Kalogera}, V. 2006, \apj, 649, 91

\bibitem
{Gai04}
{Gair}, J.~R., {Barack}, L., {Creighton}, T., {Cutler}, C., {Larson}, S.~L.,
  {Phinney}, E.~S., \& {Vallisneri}, M. 2004, Classical and Quantum Gravity,
  21, 1595

\bibitem
{Gai08}
{Gair}, J.~R. 2008, ArXiv e-prints (arXiv: 0811.0188) 

\bibitem
{Geb05}
{Gebhardt}, K., {Rich}, R.~M., \& {Ho}, L.~C. 2005, \apj, 634, 1093

\bibitem
{Geb00}
{Gebhardt}, K., {et~al.} 2000, \apjl, 539, L13

\bibitem
{Gen03a}
{Genzel}, R., {et~al.} 2003, \apj, 594, 812

\bibitem
{Ger02}
{Gerssen}, J., {van der Marel}, R.~P., {Gebhardt}, K., {Guhathakurta}, P.,
  {Peterson}, R.~C., \& {Pryor}, C. 2002, \aj, 124, 3270

\bibitem
{Ghe98}
{Ghez}, A.~M., {Klein}, B.~L., {Morris}, M., \& {Becklin}, E.~E. 1998, \apj,
  509, 678

\bibitem
{Ghe08}
{Ghez}, A.~M., {Salim}, S., {Weinberg}, N., {Lu}, J., {Do}, T., {Dunn}, J.~K.,
  {Matthews}, K., {Morris}, M., {Yelda}, S., \& {Becklin}, E.~E. 2008, in IAU
  Symposium, Vol. 248, IAU Symposium, 52--58

\bibitem
{Gur07}
{G{\"u}rkan}, M.~A., \& {Hopman}, C. 2007, \mnras, 379, 1083

\bibitem
{Hil95}
{Hils}, D., \& {Bender}, P.~L. 1995, \apjl, 445, L7

\bibitem{Hol06}
{Holley-Bockelmann}, K., \& {Sigurdsson}, S. 2006, ArXiv: astro-ph/0601520

\bibitem
{Hop05}
{Hopman}, C., \& {Alexander}, T. 2005, \apj, 629, 362

\bibitem
{Hop06}
---. 2006{\natexlab{a}}, \apj, 645, 1152

\bibitem
{Hop06b}
---. 2006{\natexlab{b}}, \apjl, 645, L133

\bibitem{Hop06c}
{Hopman}, C. 2006, astro-ph/0608460, astro-ph/0608460

\bibitem
{Hop07}
{Hopman}, C., {Freitag}, M., \& {Larson}, S.~L. 2007, \mnras, 378, 129

\bibitem
{Hop05b}
{Hopman}, C., \& {Portegies Zwart}, S. 2005, \mnras, 363, L56

\bibitem
{Hop04}
{Hopman}, C., {Portegies Zwart}, S.~F., \& {Alexander}, T. 2004, \apjl, 604,
  L101

\bibitem
{Iva02}
{Ivanov}, P.~B. 2002, \mnras, 336, 373

\bibitem
{Liu05}
{Liu}, J.-F., {Bregman}, J.~N., {Lloyd-Davies}, E., {Irwin}, J., {Espaillat},
  C., \& {Seitzer}, P. 2005, \apjl, 621, L17

\bibitem
{Men08}
{Menou}, K., {Haiman}, Z., \& {Kocsis}, B. 2008, New Astronomy Review, 51, 884

\bibitem
{Mil03}
{Miller}, J.~M., {Fabbiano}, G., {Miller}, M.~C., \& {Fabian}, A.~C. 2003,
  \apjl, 585, L37

\bibitem
{MilFabMil04}
{Miller}, J.~M., {Fabian}, A.~C., \& {Miller}, M.~C. 2004, \apjl, 614, L117

\bibitem
{Mil04b}
{Miller}, M.~C., \& {Colbert}, E.~J.~M. 2004, International Journal of Modern
  Physics D, 13, 1

\bibitem
{Mir00}
{Miralda-Escud{\' e}}, J., \& {Gould}, A. 2000, \apj, 545, 847

\bibitem
{Noy08}
{Noyola}, E., {Gebhardt}, K., \& {Bergmann}, M. 2008, \apj, 676, 1008

\bibitem
{OLe08}
{O'Leary}, R.~M., {Kocsis}, B., \& {Loeb}, A. 2008, ArXiv e-prints, (arXiv: 0807.2638) 

\bibitem
{Per07}
{Perets}, H.~B., {Hopman}, C., \& {Alexander}, T. 2007, \apj, 656, 709

\bibitem
{Rau96}
{Rauch}, K.~P., \& {Tremaine}, S. 1996, New Astronomy, 1, 149

\bibitem
{Rub06}
{Rubbo}, L.~J., {Holley-Bockelmann}, K., \& {Finn}, L.~S. 2006, \apjl, 649, L25

\bibitem
{Sch02}
{Sch{\"o}del}, R., {et~al.} 2002, \nat, 419, 694

\bibitem[{{Sesana} {et~al.}(2008){Sesana}, {Vecchio}, {Eracleous}, \&
  {Sigurdsson}}]{Ses08}
{Sesana}, A., {Vecchio}, A., {Eracleous}, M., \& {Sigurdsson}, S. 2008, \mnras,
  391, 718

\bibitem
{Sig97b}
{Sigurdsson}, S., \& {Rees}, M.~J. 1997, \mnras, 284, 318

\bibitem
{Tre02}
{Tremaine}, S., {et~al.} 2002, \apj, 574, 740

\bibitem
{Yun08}
{Yunes}, N., {Sopuerta}, C.~F., {Rubbo}, L.~J., \& {Holley-Bockelmann}, K.
  2008, \apj, 675, 604

\bibitem
{Zha09}
{Zhang}, M. 2009, ArXiv e-prints (arXiv: 0901.0301)

\end{thebibliography}

\end{document}